\documentclass{emulateapj}

\usepackage[normalem]{ulem}
\usepackage{color}
\definecolor{red}{rgb}{1.0,0.0,0.0}

\slugcomment{Submitted to ApJ June 26, 2015, Accepted October 8, 2015}

\shorttitle{HD~106906}
\shortauthors{GPI Exoplanet Survey Team}

\begin{document}

\title{Direct imaging of an asymmetric debris disk in the HD~106906 planetary system}

\author{
Paul G. Kalas\altaffilmark{1}, 
Abhijith Rajan\altaffilmark{2},
{Jason J. Wang}\altaffilmark{1},
Maxwell A. Millar-Blanchaer\altaffilmark{5}, 
Gaspard Duchene\altaffilmark{1,3,4},
Christine Chen\altaffilmark{12},
Michael P. Fitzgerald\altaffilmark{6},
Ruobing Dong \altaffilmark{1},
James R. Graham\altaffilmark{1}, 
{Jennifer Patience}\altaffilmark{2},
Bruce Macintosh \altaffilmark{10}, 
Ruth Murray-Clay \altaffilmark{8},
{Brenda Matthews}\altaffilmark{15},
{Julien Rameau}\altaffilmark{14},
{Christian Marois}\altaffilmark{15},
{Jeffrey Chilcote}\altaffilmark{5},
{Robert J. De Rosa}\altaffilmark{1},
{Ren\'e Doyon}\altaffilmark{14},
{Zachary H. Draper}\altaffilmark{15},
{Samantha Lawler}\altaffilmark{15},
{S. Mark Ammons}\altaffilmark{9},
{Pauline Arriaga}\altaffilmark{6},
{Joanna Bulger}\altaffilmark{27},
{Tara Cotten}\altaffilmark{20},
Katherine B. Follette\altaffilmark{10},
{Stephen Goodsell}\altaffilmark{11},
{Alexandra Greenbaum}\altaffilmark{23},
{Pascale Hibon}\altaffilmark{11},
{Sasha Hinkley}\altaffilmark{28},
{Li-Wei Hung}\altaffilmark{6},
{Patrick Ingraham}\altaffilmark{25},
{Quinn Konapacky}\altaffilmark{26},
{David Lafreniere}\altaffilmark{14},
{James E. Larkin}\altaffilmark{6},
{Douglas Long}\altaffilmark{12},
{J\'er\^ome Maire}\altaffilmark{5},
{Franck Marchis}\altaffilmark{16},
{Stan Metchev}\altaffilmark{29,30,31},
{Katie M. Morzinski}\altaffilmark{17},
{Eric L. Nielsen}\altaffilmark{10, 16},
{Rebecca Oppenheimer}\altaffilmark{18},
{Marshall D. Perrin}\altaffilmark{12},
{Laurent Pueyo}\altaffilmark{12},
{Fredrik T. Rantakyr\"o}\altaffilmark{11},
{Jean-Baptiste Ruffio}\altaffilmark{10},
{Leslie Saddlemyer}\altaffilmark{15},
{Dmitry Savransky}\altaffilmark{19},
{Adam C. Schneider}\altaffilmark{22},
{Anand Sivaramakrishnan}\altaffilmark{12},
{R\'emi Soummer}\altaffilmark{12},
{Inseok Song}\altaffilmark{20},
{Sandrine Thomas}\altaffilmark{25},
{Gautam Vasisht}\altaffilmark{24},
{Kimberly Ward-Duong}\altaffilmark{2},
{Sloane J. Wiktorowicz}\altaffilmark{7},
{Schuyler G. Wolff}\altaffilmark{23,12}
}
\altaffiltext{1}{Astronomy Dept., Univ. of California, Berkeley CA 94720-3411}
\altaffiltext{2}{School of Earth and Space Exploration, Arizona State Univ., PO Box 871404, Tempe, AZ 85287}
\altaffiltext{3}{Univ. Grenoble Alpes, IPAG, F-38000 Grenoble, France}
\altaffiltext{4}{CNRS, IPAG, F-38000 Grenoble, France}
\altaffiltext{5}{Dept. of Astronomy \& Astrophys., Univ. of Toronto, Toronto ON M5S 3H4, Canada}
\altaffiltext{6}{Dept. of Physics \& Astronomy, UCLA, Los Angeles, CA 90095}
\altaffiltext{7}{Dept. of Astronomy \& Astrophys., Univ. of California, Santa Cruz, CA 95064}
\altaffiltext{8}{Dept. of Physics, Univ. of California, Santa Barbara CA 93106}
\altaffiltext{9}{Lawrence Livermore National Laboratory, 7000 East Ave., Livermore, CA 94040}
\altaffiltext{10}{Kavli Institute for Particle Astrophys. \& Cosmology, Stanford Univ., Stanford, CA 94305}
\altaffiltext{11}{Gemini Observatory, Casilla 603, La Serena, Chile}
\altaffiltext{12}{Space Telescope Science Institute, 3700 San Martin Drive, Baltimore MD 21218}
\altaffiltext{13}{Dunlap Institute for Astronomy \& Astrophys., Univ. of Toronto, Toronto ON M5S 3H4, Canada}
\altaffiltext{14}{Institut de Recherche sur les Exoplanetes, Dept. de Physique, Universit\'{e} de Montr{\'e}al, Montr\'eal QC H3C 3J7, Canada}
\altaffiltext{15}{National Research Council of Canada Herzberg, 5071 West Saanich Road, Victoria, BC V9E 2E7, Canada}
\altaffiltext{16}{SETI Institute, Carl Sagan Center, 189 Bernardo Avenue, Mountain View, CA 94043}
\altaffiltext{17}{Steward Observatory, Center for Astronomical Adaptive Optics, Univ. of Arizona, 933 N. Cherry Ave., Tucson, AZ 85721}
\altaffiltext{18}{American Museum of Natural History, New York, NY 10024}
\altaffiltext{19}{Sibley School of Mechanical \& Aerospace Engineering, Cornell Univ., Ithaca NY 14853, USA}
\altaffiltext{20}{Dept. of Physics \& Astronomy, The Univ. of Georgia, Athens, GA 30602-2451, USA}
\altaffiltext{21}{NASA Ames Research Center, Moffett Field, CA 94035}
\altaffiltext{22}{Univ. of Toledo, 2801 W. Bancroft St., Toledo, OH 43606}
\altaffiltext{23}{Physics \& Astronomy Dept., Johns Hopkins Univ., Baltimore MD, 21218}
\altaffiltext{24}{NASA Jet Propulsion Laboratory, California Institute of Technology, 4800 Oak Grove Drive, Pasadena, CA 91109, USA}
\altaffiltext{25}{AURA/LSST, 950N Cherry Av, Tucson, AZ 85719}
\altaffiltext{26}{Univ. of California, San Diego, La Jolla, CA 92093}
\altaffiltext{27}{Subaru Telescope, National Astronomical Observatory of Japan, 650, North A’ohoku Place, Hilo, HI 96720}
\altaffiltext{28}{School of Physics, Univ. of Exeter, Stocker Road, Exeter, EX4 4QL, UK}
\altaffiltext{29}{Physics \& Astronomy Dept., Univ. of Western Ontario, London, ON N6A 3K7, Canada}
\altaffiltext{30}{Centre for Planetary \& Space Exploration, Univ. of Western Ontario, London, ON N6A 3K7, Canada}
\altaffiltext{31}{Physics \& Astronomy Dept., Stony Brook Univ., Stony Brook, NY 11794-3800}

\begin{abstract}
We present the first scattered light detections of the HD~106906 debris disk using Gemini/GPI in the infrared and HST/ACS in the optical.  HD~106906 is a 13~Myr old F5V star in the Sco-Cen association, with a previously detected planet-mass candidate HD~106906b projected 650~AU from the host star.  Our observations reveal a near edge-on debris disk that has a central cleared region with radius $\sim$50~AU, and an outer extent $>$500~AU.  The HST data show the outer regions are highly asymmetric, resembling the  ``needle'' morphology seen for the HD~15115 debris disk.  The planet candidate is oriented $\sim$21$\degr$ away from the position angle of the primary's debris disk, strongly suggesting non-coplanarity with the system.
We hypothesize that HD 106906b could be dynamically involved in the perturbation of the primary's disk, and investigate whether or not there is evidence for a circumplanetary dust disk or cloud that is either primordial or captured from the primary. We show that both the existing optical properties and near-infrared colors of HD 106906b are weakly consistent with this possibility, motivating future work to test for the observational signatures of dust surrounding the planet.
\end{abstract}


%%%%%%%%%%%%%%%%%%%%%%%%%%%%%%%%
\section{Introduction}
The Gemini Planet Imager Exoplanet Survey (GPIES) is targeting 600 young, nearby stars to directly detect and characterize extrasolar planets and dusty debris disks.  The general observing strategy is to obtain relatively deep ($\sim$1~hr) observations of young stars with the spectroscopic mode of the Gemini Planet Imager (GPI), and shorter snapshots using GPI's dual channel imaging polarimetry mode to detect polarized light scattered by circumstellar dust grains. The scientific motivations include quantifying the frequency and masses of Jovians from 5 to 50~AU, determining the properties of their atmospheres, and understanding their dynamical co-evolution with the planetesimals that replenish reservoirs of dust grains seen as debris disks. In particular, well-resolved debris disks typically have features such as central holes, azimuthal clumps, and vertical warps that in a single snapshot reveal key properties of each system's recent dynamical history.

Fomalhaut, HR~8799, HD~95086, and $\beta$ Pic are four prominent examples of dusty debris disks dynamically associated with directly imaged planets \citep{kalas08a, marois08a, marois10a, rameau13a, lagrange09a}, but a more recent candidate for planet-disk interactions is HD~106906 (HIP~59960; 92$\pm$6 pc; F5V; 1.5 M$_\odot$; 5.6 L$_\odot$; 13$\pm$2 Myr; \citealt{pecaut12a}). The dusty debris disk was first discovered with a {\it Spitzer} infrared survey of 25 stars comprising Lower Centaurus Crux \citep{chen05a}. Excess infrared emission in both the {\it Spitzer} MIPS 24 and 70~\micron\ bands correspond to $L_{\text{IR}} / L_\star = 1.4 \times 10^{-3}$ and blackbody radius $\sim$20~AU \citep{chen11a}. 

A comoving, substellar companion (11$\pm$2~$M_{\text{Jup}}$) was subsequently discovered at a projected separation of 7.11\arcsec (654~AU) and position angle PA = 307\fdg3 \citep{bailey14a}.  At such a large projected distance, a key question is whether or not HD~106906b originally formed like a planet in a circumstellar disk surrounding the primary and was subsequently dynamically ejected from the system, or if HD~106906b formed like a stellar companion by gravitational collapse within a common molecular cloud shared with the primary.  \citet{bailey14a} thought the former explanation was less likely because it invoked a dynamically perturbed disk observed fortuitously during the relatively brief epoch of outward planet scattering. However, the precise structure of the debris disk had not been determined by spatially resolved imaging.

Here we: 1) present new data obtained with GPI and archival data obtained with the Hubble Space Telescope (HST) that resolve the HD~106906 circumstellar disk for the first time; 2) elucidate the overall geometry of the system; 3) constrain the existence of lower mass planets within $\sim$100~AU of the primary; and 4) investigate whether or not the HD~106906b may have its own circumplanetary material. 

%%%%%%%%%%%%%%%%%%%%%%%%%%%%%%%%
\section{Observations \& Data Reduction}
HD~106906 was observed with the Gemini Planet Imager at the Gemini South 8-m telescope, Cerro Pachon, Chile on 2015, May 04. The seeing was characterized by the summit MASS/DIMM instrument as 0\farcs7/0\farcs5 full-width at half maximum (FWHM), respectively, with an atmospheric coherence time 2.4~ms (median Pachon coherence time 5~ms), indicating high low-level turbulence. Forty two 60-s frames were collected in $H$-band spectral mode with 25\fdg3 of field rotation. An additional eight 90-s exposures were collected in imaging polarimetry mode at four different waveplate orientations (0\degr, 22\fdg5, 45\degr, and 67\fdg5; two exposures per orientation) with 7\fdg1 of field rotation.
% 

%\subsection{Data Reduction}
%
The spectral and polarimetry mode data were reduced using the GPI Data Reduction Pipeline \citep{perrin15a}.  The spectral data were dark subtracted, flexure corrected and wavelength calibrated with an $H$-band Ar discharge lamp taken immediately before the science sequence.  Bad pixels were repaired in the raw 2-D data, the data were assembled into data cubes, and corrected for distortion.
The polarimetry data were dark subtracted, flexure corrected using a cross-correlation method, bad pixels repaired, corrected for the lenslet response using the Gemini calibration (GCAL) flat field lamp, and assembled into a Stokes datacube using a Mueller matrix model of the instrument. The instrumental polarization was subtracted as described in \citet{millarblanchaer15a}. 
The Stokes datacube was then converted to the radial Stokes convention ( $[I,Q,U,V] \rightarrow [I,Q_r,U_r,V]$; \citealt{schmid06a}), which places the tangentially (ortho-radial) and radially polarized intensity into the $Q_r$ image. Under the convention used here tangentially polarized intensity results in positive values in the $Q_r$ image. 

In polarimetry mode the data outside of $\sim$ 0\farcs3 is typically dominated by the photon noise of the point-spread function (PSF) halo and detector read noise \citep{perrin15a}. To improve the signal to noise, we apply a Gaussian filter with FWHM = 3 pixels (42~mas; the FWHM of Gemini's diffraction limited PSF in the $H$-band). 

The initial discovery of the disk in both the spectral and polarimetry mode data prompted us to re-examine the archival HST data obtained with the Advanced Camera for Surveys High Resolution Channel (ACS/HRC) coronagraph.  These optical data (F606W; 2 frames of 1250 seconds each) were obtained UT 2004 December 01 as part of a follow-up imaging campaign to stars with Spitzer-detected infrared excesses (GO-10330, PI Ford).  \citet{bailey14a} used these data to obtain precision astrometry on HD~106906b relative to the primary, but did not implement PSF subtraction.  The GO-10330 observing sequence included a PSF reference star (HD 103746) observed immediately after HD~106906.  Subtracting the HD~106906 PSF using the HD~103746 data, we detect the HD~106906 debris disk at a position angle that is similar to the structure discovered in the GPI data.

%%%%%%%%%%%%%%%%%%%%%%%%%%%%%%%%%%%%%%%%%%%%%%%%%%%%%%%%%%%%%%%%%%%%%%%%%%%%
\section{Analysis}
%%%%%%%%%%%%%%%%%%%%%%%%%%%%%%%%%%%%%%%%%%%%%%%%%%%%%%%%%%%%%%%%%%%%%%%%%%%%

\subsection{Disk Morphology and System Geometry}

The left panel of Figure 1 shows the GPI spectral mode $H$-band data that are PSF subtracted using \texttt{pyKLIP} \citep{wang15b}, a Python implementation of the Karhunen-Lo\`eve Image Projection (KLIP) algorithm \citep{soummer12a, pueyo15}.
The disk is clearly detected as a nearly edge-on belt of material. The most prominent feature is the slightly bowed east-west structure that does not intersect the star but instead passes north of the stellar location.  At roughly 0\farcs5--0\farcs6 radius for both extensions, the narrow  linear morphology begins to diffuse vertically toward the south, defining the ansae of the disk. The west extension in particular appears to more clearly define a fainter linear structure a comparable distance south of the star.  The canonical interpretation of such features is that preferential forward scattering makes the front (out-of-sky-plane) portion of the belt brighter than the back (into-sky-plane) portion (labeled ``backside'' in Figure 1).  

We find the position angle of the northern (out-of-sky-plane) feature by taking orthogonal cuts through the disk to find the brightest pixels between 0\farcs20 and 0\farcs46 radius.  A linear fit to these points gives PA=283\fdg7. 
The line passes 72 mas north of the star.  If we use only the higher signal-to-noise east extension to measure the position angle by a line fit, we obtain PA=284\fdg1 and the fit passes 22 mas above the star.  The uncertainty in the position angle orientation of GPI is 0\fdg13.

Though absolute photometry is unreliable due to disk self-subtraction, relative photometry shows that the east extension is 1.8 times brighter than the west extension (using the median intensity of peak pixels in the region 0\farcs20-- 0\farcs46 radius).  In addition to a brightness asymmetry, there is a related length asymmetry.  The east extension is detectable to a radius of 1\farcs06 (97~AU) whereas the west extension is detected only to a radius of 0\farcs89 (82~AU).

One of the most important measurements is to define the location of the ansae in order to characterize the size of the inner hole, to search for a stellocentric offset, and to establish the projected semimajor axis, $a$, of the structure. We note that stellocentric offsets are defined by the geometric symmetry of the disk's inner hole with respect to the star and not by the outer disk asymmetries such as the length asymmetry.  Plotting a three-pixel wide cut that intersects the star and has PA=283\fdg7, we find that the intensity in the oversubtracted region is negative (this is the region of disk self-subtraction that lies between the front and back sides of the disk), but rises above zero at 0\farcs54 radius (50 AU) and reaches a peak at 0\farcs61 radius (56 AU).  These findings are symmetric to the east and west and $a$ is most likely in this range.  We estimate the projected semi-minor axis ($b$) of the belt by measuring the median separation between the northern and southern edges of the belt in the west extension between 0\farcs19 and 0\farcs44 radius.  This gives $b$ = 0\farcs053 $\pm$ 0\farcs007.
Assuming that the structure is intrinsically circular, this estimated aspect ratio corresponds to a line of sight inclination $i$=84\fdg3--85\fdg0.  If we assume an intrinsically elliptical structure, then it may be that the semimajor axis of this intrinsic structure is pointing out of the sky plane in the region east of the star, and this partially accounts for the east-west brightness asymmetry.

The left panel of Figure 1 marks a region of the belt morphology 0\farcs27--0\farcs39 west of the star where the belt structure warps to the south by $\sim$ 0\farcs03.  This feature is not detected $\sim$0\farcs3 east of the star. The existence of this warp is tentative and requires confirmation.  However, when taken together (the potential warp, and the asymmetries in brightness and length) the general conclusion is that the HD~106906 belt has a more complicated, morphologically disturbed structure than the more azimuthally symmetric HR~4796A debris belt \citep{schneider09a}.  

The middle panel of Figure 1 shows the result of using a different PSF subtraction technique on the same data.  Here we attempt to minimize the self-subtraction of the disk that is evident with the KLIP PSF subtraction (left panel of Figure 1).  In every image we block a rectangular region 20 pixels (0\farcs28) wide and with the long axis oriented and centered along the disk midplane on each side of the star.  The background PSF is sampled outside of the masked region and a low-order polynomial fit is used to generate interpolated PSF values over the masked rectangular region; this PSF is further smoothed using an 11 pixel (0\farcs15) running median filter to reduce edge effects and focus on subtracting the low spatial frequency structure of the PSF.  A similar PSF subtraction technique was used for the GPI study of the HR 4796A disk (\S\ 5.4 in \citealt{perrin15a}).  

The interpolated PSF subtraction has a larger inner working angle (0\farcs36) than the KLIP scheme, but mitigating the self-subtraction means that the disk flux and the vertical morphology is better preserved.  Thus, the inner region with the tentative warp is too close to the star to be imaged using this technique.  However, this PSF subtraction confirms the greater radial extent of the east ansae, and the morphology of the west ansae that shows the structure curves southward at 0\farcs54 radius to define the ``backside'' of the disk.  As with the KLIP-processed image, orthogonal cuts through the disk to find the brightest pixels between 0\farcs35 and 0\farcs46 radius for both sides of the disks result in a line that gives PA = 283\fdg7 and passes 51~mas north of the star.  These results are consistent with the KLIP processed image.

With self-subtraction mitigated, the peak pixel values are greater in the middle panel of Fig. 1 compared to the left panel by a factor of $\sim$17 at 0\farcs37 radius, and a factor of 2-3 at 0\farcs70 radius.  
Though the east-west brightness asymmetry is again evident close to the star in the interpolated PSF subtraction image, it is not as prominent and does not extend over a larger radial region.
In the radial region 0\farcs37--0\farcs38 the east extension is 20--25\% brighter than the west extension.  Between 0\farcs38 radius and 0\farcs69 radius  the disk brightness is symmetric between the east and west sides, but at 0\farcs69 radius the west extension is truncated, whereas the east extension is detected to 0\farcs83 radius.  Thus the length asymmetry seen in the KLIP image is confirmed.  The brightness profile of the brightest pixels along the disk as a function of radius in the region 0\farcs37--0\farcs69 can be fit by power laws with exponents -3.0 and -3.3 for the east and west extensions, respectively.

The interpretation that there is a near-edge-on belt that comes out of the sky plane north of the star is further supported by the polarization intensity image ($Q_r$; Figure 1, right panel). In polarized intensity the disk only appears above the midplane to the north of the star and is detected roughly symetrically to $\sim$ 0\farcs9 radius to the east and the west. The polarization intensity image also shows less east-west brightness asymmetry, consistent with the total intensity image using the interpolated PSF subtraction scheme (middle panel).  In the 0\farcs37--0\farcs38 radial region, the east extension is $\sim$ 20\% brighter than the the west extension.  However, the image does not show the radial truncation of the west extension relative to the east extension, implying that beyond 0\farcs69 radius the fractional linear polarization is greater in the west than the east. 

To estimate the position angle of the disk in the polarized image we again find the maximum pixel in each vertical column using only pixels within 0\farcs9 of the central star and ignoring columns that contain the area masked out by the focal plane mask (less than $\sim$ 0\farcs12). The $x$- and $y$-pixel positions of the maximum pixel were then fit by a straight line, giving PA = $284.1\degr$  
The orthogonal distance between the line fit and the stellar position is 51 mas, which again is consistent with the previous measurements.

Detection of polarized emission with GPI does not rely on angular differential imaging (ADI) and pyKLIP to subtract the PSF.  With ADI, azimuthally extended structures near the star, such as debris disks, will self-subtract.  For example, an edge-on disk that has some intrinsic vertical width will appear artificially narrow after ADI processing \citep{Milli2012, esposito14a}, but polarized intensity images are not susceptible to this effect. Indeed the structure of the HD 106906 disk in polarized light appears more vertically extended than in images processed with pyKLIP, which we interpret to be more representative of the true width of the disk in projection. We estimate the disk width by fitting a Lorentzian function to the vertical profile (i.e. perpendicular to the disk midplane) averaged between $0.25\arcsec$-$0.35\arcsec$ on either side of the disk. We find the FWHM of both the east and west sides to be $\sim0.13\arcsec$, well above the resolution of the images.

We defer modeling of the HD~106906 grain properties and disk structure to future work.  However, here we briefly consider the idea that the east-west brightness asymmetry is due to an intrinsically elliptical disk with a stellocentric offset. To test this hypothesis, we build a toy model consisting of an azimuthally uniform narrow ring (5\,AU width, 75\,AU inner radius) with the star offset in the sky plane by 15\,AU (0.16'') from the ring geometric center, effectively simulating an eccentricity of 0.2. The ring is assumed to contain an optically thin amount of dust grains whose composition is astronomical silicates \citep{draine_lee84} and whose grain size distribution ranges from 1.5\,$\mu$m to 1\,mm with a power law distribution $N(a) \propto a^{-3.5}$. We use MCFOST \citep{pinte06, pinte09} to compute scattered lights for all Stokes parameters at 1.65\,$\mu$m. The resulting images are shown in Figure~2 after convolution with a 3-pixel FWHM Gaussian kernel to mimic the GPI PSF. This model shows that the east-west brightness asymmetry could be due to the proposed geometry, but a stellocentric offset should also be observed. The model also predicts a brightness asymmetry in the polarized intensity image, but the noise level in the observed map (Fig. 1) is insufficient to determine whether it is indeed present.

This toy model illustrates how a standard dust composition could account for the total and polarized intensity appearance of HD 106906b. Future data sets and models will need to explore the parameter space more thoroughly to self-consistently fit the observations. Our current model simply demonstrates that an azimuthally uniform disk can display a left-right brightness asymmetry, but this also requires a stellocentric offset, which is not currently observed.  Therefore it is likely that the left-right asymmetry is due to an azimuthally and/or radially asymmetric distribution of dust in the 10 -- 100 AU radial region surrounding HD 106906.

The optical HST/ACS data (Figure 3) show what appears to be a near edge-on disk midplane extending to nearly 6\farcs0 (550~AU) radius to the west, but the image lacks a corresponding feature 180\degr\ to the east.  Instead, the eastern component consists of a fan-shaped region of diffuse nebulosity detected to $\sim$ 4\farcs0 (370~AU) radius.  Therefore, on these much larger spatial scales, the length asymmetry observed in the GPI data is reversed.  In the HST data the apparent disk midplane to the west is roughly consistent with the 284\degr\ position angle detected with GPI in $H$-band.  Between 2\farcs75 and 5\farcs15 radius we measure the location and intensity of the peak value in 0\farcs2 wide cuts perpendicular to the midplane. The peak pixel surface brightness at 2\farcs75 is 19.4 mag arcsec$^{-2}$ (for F606W in the Vega magnitude system without aperture corrections) and drops to 21.9 mag arcsec$^{-2}$ at 5\farcs15 radius as a power law function with radius that has exponent of approximately -3.6.  

A linear fit to the 12 measurements of intensity peak location gives PA = 286\fdg3 and the fit extrapolated toward the star passes 250~mas north of the star. This is consistent with the inferences from the GPI image that the disk is not exactly edge-on, but rather has a $\sim$ 85\degr\ inclination.  
The PA is $\sim$2\degr\ greater than that measured for the $H$ band images, and the offset between a line fit and the stellar location is five times larger.  There are at least four explanations for such apparent discrepancies:  (1) The HST image is contaminated by residual radial noise features that can be mitigated with follow-up imaging to improve the PSF subtraction and signal-to-noise, (2) optical and infrared images probe different grain size regimes, (3) the radial regions probed are very different, as shown by the scale bars in Figures 1 and 3, and (4) there are distinct differences in the morphological asymmetries in these radial regions $-$ the east extension of the HST detected disk may indeed bend over large radial scales toward larger PA \citep[as seen, for instance, with the HD 32297 debris disk;][]{kalas05a}.

Figure 3 also demonstrates the $\sim$21\degr\ difference in position angle between the ACS detected disk and the low mass secondary companion.  This angle was previously unknown, nor was it known that cold dust surrounding the primary extends to $>$ 400~AU.  For the assumption that HD~106906b is coplanar with the belt, and given its location $\sim$2\farcs5 north of the belt midplane 
and a belt inclination of $\sim$85\degr, its line of sight position is $\sim$2600~AU out of the sky plane toward the observer.  Since the primary's disk is vertically disturbed, we could alternately assume that the orbital plane of HD~106906b is in fact misaligned with the primary's belt.  For example, the 99~Her system has a circumbinary debris disk that is misaligned by $>$ 30\degr\ relative to the orbital plane of the central binary \citep{kennedy12a}.

%%%%%%%%%%%%%%%%%%%%%%%%%%%%%%%%%%%%%%%%%%%%%%%%%%%%%%%%%%%%%%%%%%%%%%%%%%%%
\subsection{Search for Additional Planets with GPI}

Figure~\ref{fig:Contrast} translates our point source detection limits with GPI to planet mass detectability under a variety of model assumptions. The important point is that no planet as massive as HD~106906b (11~M$_J$) is detected in the GPI field, with an inner working radius of 0\farcs2 (18.4~AU). This is relevant because if HD~106906b formed in a circumstellar disk around the primary and was subsequently ejected to large radii by planet-planet scattering, a perturber with comparable or greater mass  might still reside in the system.  Unfortunately, our search with an 18.4~AU projected inner radius is not exhaustive;  for example, a  $\beta$ Pic b analog with a $\sim$ 9~AU semimajor axis and low eccentricity would remain hidden around HD~106906b with the current GPI data (though planned non-redundant aperture masking with GPI can probe closer to the star).

\subsection{Optical Photometry of HD~106906b with HST}

Our recovery of HD~106906b with HST/ACS validates the \citet{bailey14a} discovery in these data. Overall, we confirm their astrometric measurements, but can refine their F606W photometry, which they give as ``[F606W] = 24.27 $\pm$ 0.03 mag''.  In our version we restrict our measurement to the first 1250-second exposure ({\tt j917711lkq\_drz.fits}) because there are no cosmic ray hits within the boundary of the first Airy ring.  We use the same PSF subtraction as displayed for Fig. 3, but the data are not rotated to north to avoid interpolation artifacts.  We measure photometry within 0\farcs2 radius and our estimate for the sky background value is the median value of pixels contained in an annulus between 0\farcs200 and 0\farcs375 radius.  The photometry within 0\farcs2 gives 1.934 electrons/second.  We use the information provided by \citet{chiaberge09a} to adopt a 0.009~mag charge-transfer efficiency (CTE) correction.  
For the aperture correction, \citet{sirianni05a} give 0.180~mag at 0.600~$\mu$m to correct from an 0\farcs20 aperture to a 5\farcs5 radius aperture.  However,they recommend that the encircled energy profiles of stars in the observations at hand are used due to various effects such as differences in focus.  For the five brightest field stars we empirically determined the aperture correction from 0.2$''$ to 0.5$''$.  This gives a median value of 0.212$\pm$0.005~mag.  Then we used \citet{sirianni05a} to add the published aperture correction from 0.5$''$ to 5.5$''$ radius, which is 0.089~mag.  Thus the average aperture correction is 0.301~mag.  We arrive at the final photometric measurements for HD~106906b, which are CTE and aperture corrected:  VEGAMAG = 24.07 mag, STMAG = 24.31 mag, ABMAG = 24.15 mag (or 0.800 $\mu$Jy, assuming a F606W zeropoint of 3630~Jy).  The 1-$\sigma$ uncertainty in deriving an aperture correction is 0.005 mag.  However, PSF subtracted images have residual background fluctuations that dominate the photometric noise.  To empirically estimate the photometric measurement uncertainty for a source this faint, we inserted 13 copies of a TinyTim PSF (appropriately scaled to the flux of HD~106906b) into the regions free of cosmic-ray hits within 2$''$ of the location of HD~106906b.  We then performed aperture photometry using exactly the same technique as for HD~106906b, and determined a 1-$\sigma$ photometric uncertainty of 0.14 mag. 

%%%%%%%%%%%%%%%%%%%%%%%%%%%%%%%%
\section{Discussion}
%%%%%%%%%%%%%%%%%%%%%%%%%%%%%%%%
\subsection{Dynamical Paradigms:  Disk-Planet Interaction?}
The detections of a highly asymmetric outer disk, a moderately asymmetric inner disk, and a distant substellar companion to HD~106906 offset by $\sim$21$\degr$ raises new questions about the system's dynamical history.  There are at least three scenarios to consider based on the assumed formation site for HD 106906b and the body responsible for perturbing the outer disk:
\begin{enumerate}
\item HD~106906b formed in the natal circumstellar disk near the primary, and it was subsequently ejected to $\geq$650~AU via planet-planet scattering or by some other instability \citep[e.g.][]{weidenschilling96a, rasio96a, ford01a, veras09a}.  HD 106906b currently has an eccentric orbit and produces the significant outer disk asymmetries discovered with HST.  However, this hypothetical scenario has several issues that require further observational and theoretical testing:

\textbf{(a)} The assumption that HD~106906b formed in a disk near the star invokes a complex dynamical history involving additional massive perturbers.  With planet-planet scattering the perturber would have to be nearly as massive as HD~106906b \citep[e.g.][]{chatterjee08a, juric08a}.
Moreover, the mild asymmetry of the inner disk compared to the strong asymmetry of the outer disk suggests that the periastron of HD~106906b probably resides beyond $\sim$100 AU at the current epoch. Its dynamical history therefore begins with the planet's formation close to the star in a disk, its eccentricity subsequently increases through interactions with other massive planets, and finally its periastron is increased by interactions with other cluster members at the planet's apastron \citep[e.g.][]{scally01a, malmberg11a, vincke15a}.  The wider field should therefore be searched for other candidate perturbers, and multi-epoch imaging, astrometry and radial-velocity need to tighten the constraints on a second massive planet that may be hidden behind GPI's coronagraphic spot at the current epoch.

\textbf{(b)}  If HD 106906b recently exited the inner system by some form of dynamical upheaval involving gas giant planets, then the morphology of the inner disk should probably appear more strongly asymmetric.  However, additional dynamical modeling is required to explore the validity of this concern.

\textbf{(c)}  \citet{jilkova15a} have studied the possible disk morphologies resulting from repeated encounters between HD 106906b on an eccentric orbit and the debris disk around the primary.  Very strong asymmetries are possible, but the specific observed asymmetries remain to be tested. A larger search of the orbital parameter space needs to be conducted to establish if the vertical disturbance on one side can coexist with the radially extended feature on the other side. The general concern is that over the timescales required to vertically excite the disk, the flatter, radially extended side of the disk will precess and lose its prominence as a one-sided feature.

\item HD~106906b formed like a star far from the primary, as favored by \citet{bailey14a}. If HD 106906b is bound to the primary and perturbs its dust disk, then we are again led to the area of concern described in \textbf{1(c)}.  If HD 106906b is unbound, then could a single close approach perturb the disk? \citet{larwood01a} demonstrate that a stellar flyby can briefly result in an extremely asymmetric disk with one side that is flat and radially extended and an opposing side that is vertically extended and radially truncated.  However, to produce the observed vertical excitation over a significant portion of the disk in a flyby event, stellar mass ratios of a few tenths are required.  The planet-star mass ratio for the HD 106906 system is $\sim$0.01, which is too small.  Therefore, HD 106906b as an unbound object would not directly create the large-scale disk asymmetry.  

\item Scenarios 1 and 2 attempt to causally associate the properties of HD 106906b and the primary's asymmetric disk.  A third hypothetical scenario is that the planet and disk are independent.  The assumptions adopted for the formation site and evolution of HD 106906b are irrelevant for the outer disk asymmetry.  Instead, the observed outer disk asymmetry was recently created by a stellar flyby.  As noted in \textbf{1(a)}, a stellar-mass cluster member at an earlier epoch may have interacted strongly with the disk surrounding HD 106906. HD 106906b may have also been perturbed, but in this scenario there is no direct relationship between the planet and the primary's disk.  
\end{enumerate}

Future work is clearly essential to distinguish the relative merits of these three scenarios.  Given the data at hand, a possible consequence of HD 106906b interacting with the primary's debris disk is the capture of dust.  In this case, HD 106906b may have an IR excess and exhibit unusual reddening.  Moreover, the optical HST flux may be anomalous due to light reflected from a larger circumplanetary ring or cloud, as has been hypothesized to explain the anomalous optical flux of Fomalhaut b \citep{kalas08a}.
Such a hypothetical circumplanetary dust disk or dust shroud may appear extended in high angular resolution data.
Of course, finding a dust disk surrounding HD 106906b would not be definitive proof that the planet and disk are interacting because the origin of circumplanetary dust could be primordial.  For example, the diversity of infrared colors exhibited by substellar objects has been attributed to dust disks \citep[e.g.][]{mohanty07a}, among other explanations, such as dust in the atmospheres and non-equilibrium chemistry \citep[e.g.][]{barman11a}.  Nevertheless, the proximity of HD 106906b to the primary's debris disk motivated us to test the existing data for evidence of circumplanetary material.

\subsection{Testing for a circumplanetary disk}

To search for evidence of a circumplanetary disk, we conducted three experiments: (1) measure the radial profile of HD~106906b in the HST image to determine if the object is extended, (2) test whether or not scattered light could account for the optical flux, and (3) compare the colors of  HD 106906b to both model predictions and an empirical sample of other bound, substellar objects with similar ages and spectral types. 

\subsubsection{Radial profile of HD~106906b}

With the stable PSF delivered by HST and the presence of numerous additional point sources in the ACS image, the HD~106906b radial profile can be tested for extended nebulosity.  If HD~106906b is an 11 M$_J$ object 650~AU from a 1.5 M$_\odot$ star, the Hill sphere has radius 86~AU (0.93$''$). 
If the dust was captured when HD 106906b was located closer to the star, the Hill radius would be smaller (e.g. the planet at 100 AU has a Hill radius of 13 AU or 0.14$''$).  Therefore, a debris cloud surrounding HD~106906b could be resolved.

To test for extended nebulosity in the F606W data, we measured the radial profiles of HD~106906b and 11 other point sources in the field.  Figure 5 demonstrates that the PSF core of HD~106906b is consistent with the other 11 sources, but the PSF wing is anomalously bright between 0.10\arcsec\ and 0.15\arcsec\ radius (9--14 AU).  Specifically, in this radial region the HD~106906b PSF has 26\% more summed light than the summed light of the average PSF from field stars (all PSF peaks are normalized to unity). Or, including the cores of the PSF, the summed light from 0\arcsec\ to 0.15\arcsec is 1.6\% greater.  Therefore, the extra PSF halo brightens the optical magnitude of HD~106906b by 0.017~mag.  

The PSFs are also distinguished by the radius at which the first Airy ring peaks.  For HD~106906b the peak is at 4.44 pixels radius (111 mas or 10.2 AU), whereas for the 11 field stars the Airy ring peaks at a median value of 3.69$\pm$0.23 pixels. To estimate the uncertainty for the value of the HD 106906b Airy ring maximum, we turn to the 13 artificial point sources that were inserted into the data for the purpose of determining the photometric uncertainty (Section 3.3).  As discussed below, the TinyTim PSF does not exactly represent the astrophysical PSF of HD 106906b, but we can nevertheless use it to quantify how the measurements of the Airy ring peaks are influenced by noise at the 13 different insertion points near HD 106906b.  This experiment shows that the Airy peak measurement on a source as faint as HD 106906b has $\sigma$ = 0.14 pixel.  Adding the two uncertainties in quadrature, the difference between the planet and the field star Airy ring peaks is 0.71$\pm$0.27 pixels.  

To test whether or not the PSF shape is due to the extreme red color of HD~106906b, we examined the TinyTim \citep{krist11a} calculations of PSF structure for HST/ACS/HRC coronagraphic observations in F606W.  We find that the first Airy ring for an A0V star, an M3V star, and a 1000 K blackbody peaks at 3.52 pixels (0.088$''$), 3.68 pixels, and 3.68 pixels radius, respectively.  The maximum flux level of the 1000~K Airy ring is 5\% lower than the A0V Airy ring.  HD~106906b, on the other hand, has significantly more flux in the first Airy ring compared to the comparison objects in the field, and the ring peaks at a greater radius, as shown above.  The TinyTim models for PSF structure therefore do not account for the extended HD~106906b PSF size.

The experiments above give tentative evidence for a slightly resolved structure surrounding HD 106906b.  What we were identifying as the peak of an Airy ring around HD 106906b should instead be termed a shoulder on top of the intrinsic Airy ring.  Nevertheless, it is critically important to observe HD 106906b to greater depth and with different instrumentation to understand if spurious noise and/or a distant background object could account for the shoulder detected in the ACS data.

\subsubsection{Origin of optical flux for HD~106906b}

We also examined whether or not the measured F606W flux is higher than expected from the calculated in-band integrated flux of model atmospheres, matching the published effective temperature, age and mass of the companion from \citet{bailey14a}.  A similar exercise was conducted with the HST optical discovery of Fomalhaut b, which was found to have flux two orders of magnitude greater than that predicted by models \citep{kalas08a}.  For HD~106906b, the BT-Dusty and BT-Settl models (scaled to the $J$-band data) predict F606W apparent VEGAMAG magnitudes 24.64 ($5.71\cdot10^{-7}$ Jy) and 25.68 ($2.23\cdot10^{-7}$ Jy), respectively.  Our measured F606W value of 24.07 mag is 0.57 mag and 1.61 mag brighter, respectively.  The combined uncertainty of the F606W flux (0.14 mag) and the $J$-band data (0.3 mag) is $\sigma=$0.33 mag, which means that the observed optical flux is  $1.7~\sigma$ and $4.9~\sigma$ greater than the respective theoretical predictions.  This is certainly not as large a discrepancy as in the case of Fomalhaut b, but it is consistent with the hypothesis that captured material would add reflected light to the intrinsic flux from the planet.  The two important caveats, as noted in Section 4.1, are that the atmosphere models are uncertain, and there is an intrinsic astrophysical diversity in the colors of low-mass objects.

Hypothetically, we find that all of the optical light from HD 106906b could in fact arise from scattered starlight.  The stellar flux received at Earth (assuming $D$=92 pc, L$_\star = 2.143\cdot10^{27}$ W) is $f_\star = 2.116\cdot10^{-11}$ W m$^{-2}$.  For a star-planet separation of $d$=1000 AU, the stellar flux received at the planet is $7.58\cdot10^{-3}$ W m$^{-2}$ (Fomalhaut b was 1.7 W m$^{-2}$).  The reflected light will depend on several factors such as the geometry of the system, the total scattering surface  ($\Sigma$) from the planet and its dust cloud or ring, and a scattering efficiency, $Q_s$, such as the product of the geometric albedo and phase function at a given phase.  For a circumplanetary ring such as Saturn's main rings, the scattering geometry is important, but for more radially extended dust distributions we can reasonably assume an optically thin and roughly spherical dust cloud. Therefore all the grains are illuminated and it does not matter how the planet is oriented relative to the incident light and the observer.  We can write the flux received at earth as:  

\begin{eqnarray*}
f_p &=& {7.58\cdot10^{-3}~\Sigma~Q_s \over 4~\pi~D^2}\nonumber\\
&=& {7.48 \cdot 10^{-41}~\Sigma~Q_s~[{\mbox W~ \mbox m}^{-2}]}
\end{eqnarray*}
We can rewrite this as a contrast in apparent magnitude between the planet's reflected light and the star:
\begin{eqnarray*}
m_p - m_\star &=& m_p - 7.81~\mbox{mag}
=-2.5 ~\mbox{log} ~(f_p / f_\star) \nonumber\\
m_p &=& -2.5~\mbox{log} (\Sigma \cdot Q_s) + 81.44~\mbox{mag}
\end{eqnarray*}

For the sake of argument, we assume the albedo and phase function average to $Q_s$ = 0.1 and then ask how large does $\Sigma$ have to be in order to satisfy our F606W magnitude of $m_p=24.07$ mag?  In this case  $\Sigma = 10^{24}$ m$^2$ which in the geometry of a large circular disk projected onto the sky has radius r = 1.669$\cdot10^{12}$ m = 11 AU = 0.12$''$.  Observationally, this value is similar to the radial extent of the anomalous PSF shown in Figure 5 and the possible range of Hill radii given in Section 4.2.1.  If this projected surface area is due to dust grains with radius 5 $\mu$m and density 2000 kg m$^3$, then the total mass is $\sim10^{22}$ kg (i.e. similar to Pluto). Therefore the hypothetical size and mass of the dust cloud do not violate any observational or theoretical constraint.  Some or all of the optical light could arise from a circumplanetary dust cloud scattering stellar light.

%%%%%%%%%%%%%%%%%%%%%%%%%%%%%%%%%%%%%%%%%%%%
\subsubsection{Infrared Colors of HD~106906b}

A circumplanetary dust disk or cloud would be very cold due to the relatively low luminosity of the planet and the large distance from the host star.  We studied the 2MASS, Spitzer, Herschel and ALMA data and do not detect a source at the location of HD~106906b, which is expected given its extremely low luminosity (2.3$\times10^{-4}$ L$_{\odot}$; Bailey et al. 2014).  For example, given the noise properties of the Sco-Cen Spitzer observations in aggregate, uncontaminated observations place a 3-$\sigma$ limit on the 24~$\mu$m emission of $\sim$0.3~mJy (this is a best-case limit because of source confusion.) If we assume that the peak for the emergent thermal emission arises at 24~$\mu$m, then we can approximate $L_{IR} < 9.7\times10^{-6}$  L$_{\odot}$ assuming the same heliocentric distance as the primary (92~pc). Given the instrument configuration and the integration time, we can only place an upper limit on $L_{IR} / L_{planet} <$ 0.042. 

We also investigated whether or not the NIR photometry for HD~106906b published in \citet{bailey14a} is anomalous empirically (relative to several comparison objects), and theoretically (when compared against two different atmospheric model predictions). 
The infrared photometry is compiled in Table~\ref{tab:Photometry} along with the photometry for a set of seven low mass companions with similar masses and ages \citep{kraus14a, kraus15a, bailey13a, lagrange09a, patience12a, delorme13}. The comparison sample is plotted on a color-magnitude diagram in Figure~\ref{fig:CMD} (left) and all the young (2 -- 30 Myr) imaged companions with spectral types of L0 to L4, similar to HD~106906b. The young, planetary mass companions to 2M1207 and HR8799 are not included due to their considerably cooler temperatures and later spectral types. The distribution of the comparison, young low mass companions in estimated mass and age is given in Figure~\ref{fig:CMD} (right) and shows that the sample can serve as an analogous comparison sample.

Two evolutionary models - BT-Dusty \citep[][]{allard01a} and BT-Settl \citep[][]{allard12a} - were used to estimate the photospheric colors; both grids provide photometry values for objects that span the full range of ages and masses covered by the low mass companions. The grid points were interpolated with a power law to estimate magnitudes at the specific ages of the target using a model mass consistent with each target. The model photospheric colors for each target and each model are listed in Table~\ref{tab:ModelComp} with the measured colors for HD~106906b and the comparison sample. This approach to inferring the presence of a disk from photometry is similar to previous studies \citep[e.g.][]{bailey13a}.

Several of the comparison objects have previously reported evidence for disks. The most substantial disk has been detected around FW~Tau b with ALMA continuum emission \citep{kraus15a} and accretion signatures \citep{bowler14a}. Both GSC 6214-210B and 1RXS 1609-2105B exhibit excess emission \citep{bailey13a, wu15a}, and GSC 6214-210B also shows both H$\alpha$ and Pa$\beta$ emission from accretion signatures \citep{bowler11a, zhou14a, bowler14a}. In Figure~\ref{fig:colorComp}, the difference in observed and model colors is plotted as a function of age, which is also expected to correlate with surface gravity. FW~Tau b,the object with the strongest evidence for a disk, stands out as the reddest object, HD~106906b has the second largest offset from the model photospheres. The HD~106906b color excess is larger than the two other comparison objects with reported evidence of disks - GSC 6214-210B and 1RXS 1609-2105B. 

Due to the 0.3 mag uncertainty on the $J$-band photometry, the significance of the red excess is limited, but the results suggest the possibility of the presence of a circumplanetary dust around HD~106906b. Based on the combination of evidence from the IR color, \emph{HST} optical radial profile, and the optical flux level, we conclude that there may be a disk of material that was either captured in an encounter with the primary star's disk, or retained from the time of formation of the planetary mass companion. Additional observations are required to clarify these tentative conclusions about the environment surrounding HD~106906b.

%%%%%%%%%%%%%%%%%%%%%%%%%%%%%%%%%%%%%%%%%%%%%%%%%
\subsection{Comparison to HD~15115 and Fomalhaut b}

The HD~15115 debris disk was the first in what seems to be a class of debris disks that are so extreme in their disturbed morphology, they resemble a ``needle'' in the near edge-on view over 10$^2$~AU scales \citep{kalas07a}.  On scales of $\sim$10~AU, \citet{mazoyer14a} discovered that HD 15115 has a more symmetric inner hole, essentially representing the ``eye of the needle.''  As with our GPI image of HD~106906b (Figure 1), the eye of HD 15115 has a northern edge that is significantly brighter due to preferential forward scattering and a $\sim87\degr$ line of sight inclination. 
\citet{kalas07a} suggested that a nearby M dwarf may have perturbed the HD 15115 disk, though this scenario was found unlikely by \citet{debes08a}, and therefore the origin of the extreme asymmetry for HD 15115 remains an open question. The discovery of a needle-like debris disk around HD~106906 represents a fresh opportunity to investigate the origin of such structure.

The question of how HD~106906b obtained an apparent position outside of the primary's debris disk invites comparisons to the Fomalhaut system.  Fomalhaut b is currently located very near the inner edge of the debris disk, but its highly eccentric orbit will place it  beyond the outer edge in the future \citep{kalas13a}.  Its low mass ($\lesssim 1 M_J$; \citealt{janson12a}) means that the prominent 140 AU dust belt may survive many planet crossings whereas the high mass of HD~106906b tends to argue that it did not recently encounter the inner disk of the system.  In both cases, significant future work is necessary to answer the fundamental question of whether or not the planet is coplanar with the disk.  However, in both cases the question is raised on whether or not the planet has acquired circumplanetary material due to the possible interactions with the debris disk.  For Fomalhaut b the evidence for circumplanetary material rests on the anomalously high optical flux, whereas for HD~106906b the evidence is based on the possible infrared excess, the brighter optical flux than the model predictions, and the extended shape of the optical PSF compared to all other field stars.  Fomalhaut b may also be extended in the optical \citep{galicher13a}, but this result is also tentative given that the extended morphology is detected in only one bandpass (F814W).

\section{Conclusions}
New observations with GPI in $H$-band and analysis of archival coronagraphic HST data in the optical resolve the dusty debris disk surrounding the F5V star HD 106906 in scattered light.  We find:

\begin{enumerate}
\item The total intensity image obtained with GPI over a $\sim$1$''$ radius ($\sim$92 AU) field of view shows the dust disk has a central cleared region with radius $\sim$50 AU and inclination $\sim$85$\degr$.

\item The GPI images show several asymmetries:  (a) the east disk extension is detected to a greater radius than the west extension,  (b) the east extension is 20\% brighter than the west extension, and (c) a possible vertical warp may exist in the west extension at 0.3$''$ radius.

\item The complementary $H$-band polarization detection with GPI shows that the polarization intensity follows the east-west brightness asymmetry observed in total intensity, but these data do not have the radial truncation of the west extension.  This suggests that the polarization fraction increases to the west.  

\item The optical HST data on larger scales show a highly asymmetric morphology in the class of ``needle-like'' disks.  The projected semi-major axis of the west extension is misaligned with the candidate planet by $\sim$21$\degr$, suggesting that either HD~106906b did not form in a circumstellar disk surrounding the primary, or that the system is in a state of dynamical upheaval resembling the Fomalhaut system.

\item We outline three dynamical scenarios that require significant follow-up observational and theoretical testing.  In two of the scenarios, the planet is causally linked to the observed large scale disk asymmetry. We speculate that the planet could have captured material during encounters with the disk.

\item We search the existing data for evidence of circumplanetary material.  We find that the optical PSF of HD 106906b is radially extended compared to 11 comparison point sources in the HST data.  Analysis of the near-infrared photometry and models shows that HD 106906b is redder than a comparison sample of sub-stellar companions, except for FW Tau b, which has strong evidence for a circumplanetary disk.  
\end{enumerate}

These initial findings regarding a possible circumplanetary disk and the hypothesis of captured material as the origin require significant follow-up work for validation.  We reserved a thorough modeling of dust properties and debris disk structure for a future study in which we anticipate including higher signal-to-noise polarization data to be made available with GPI.  Also, there is a significant region between 100 AU (the outer edge of the GPI field) and 250 AU (the inner edge of the HST field) that has yet to be imaged, and this zone probably contains the transition between the weak asymmetries of the inner disk and the strong asymmetries of the outer disk. Analysis of the disk morphology here may help constrain the possible periastron distance of HD~106906b.  More generally, future work should search for the existence of other candidate perturbers both closer to the primary than our observations permit and in the wider field. Higher quality photometry in the infrared, particularly $J$-band, and measurements of accretion sensitive lines such as H$\alpha$ will provide a better indication of the presence of circumplanetary dust. HST follow-up imaging can ascertain whether or not the optical PSF of HD~106906b is indeed extended, and higher signal-to-noise would provide a more stringent test of the radial and azimuthal structure of any nebulosity surrounding the planet.

\acknowledgments
{\bf Acknowledgements:}  The Gemini Observatory is operated by the AURA under a cooperative agreement with the NSF on behalf of the Gemini partnership: the National Science Foundation (United States), the National Research
Council (Canada), CONICYT (Chile), the Australian Research Council (Australia),
Minist\'erio da Ci\'encia, Tecnologia e Inova\c{c}\=ao (Brazil), and Ministerio de Ciencia,
Tecnolog\'ia e Innovaci\'on Productiva (Argentina).
This research was supported in part by NASA cooperative agreements NNX15AD95G, NNX14AJ80G, and NNX11AD21G, NSF AST-0909188, AST-1411868 and AST-1413718, and the University of California LFRP-118057. This work benefited from NASA's Nexus for Exoplanet System Science (NExSS) research coordination network sponsored by NASA's Science Mission Directorate.  We thank an anonymous referee for comments that improved our manuscript.

Facilities: \facility{Gemini:South(GPI)}

\begin{figure*}
\centering
\includegraphics[width=7in]{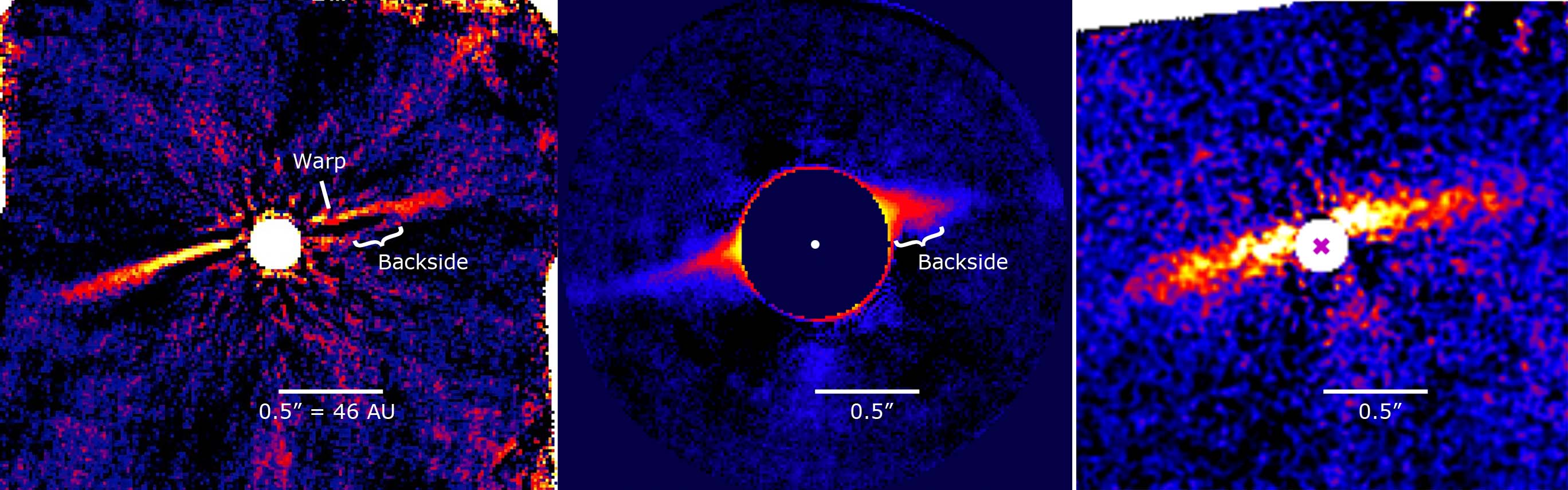}
\caption{{\bf LEFT:} GPI $H$-band spectral data cube with KLIP PSF subtraction based on angular differential imaging (north up, east left). The most prominent linear feature from disk scattered light does not intersect the position of the star.  
The overall structure appears to be a ring inclined $\sim$5\degr from edge-on.  The northern side could be brighter because it points out of the sky plane and dust grains have preferential forward scattering.  We mark the locations of the warp and the backside of the belt, as discussed in the text. {\bf MIDDLE:} The same $H$-band data as in the left panel, but the PSF subtraction is achieved by reconstructing the PSF from image flux values outside of the region encompassed by the disk.  The circular masked region has 0\farcs36 radius and the warp is blocked, but the structure we interpret as the backside of the disk is confirmed (note that the white bracket is registered to the same location in both the left and middle figure panels).
{\bf RIGHT:}  $H$-band GPI tangentially polarization intensity ($Q_r$) image. The disk appears almost exclusively to the north side of the midplane and no 'backside' is detected to the south. 
}
\label{fig:GPI}
\end{figure*}

\begin{figure*}
\centering
\includegraphics[width=5in]{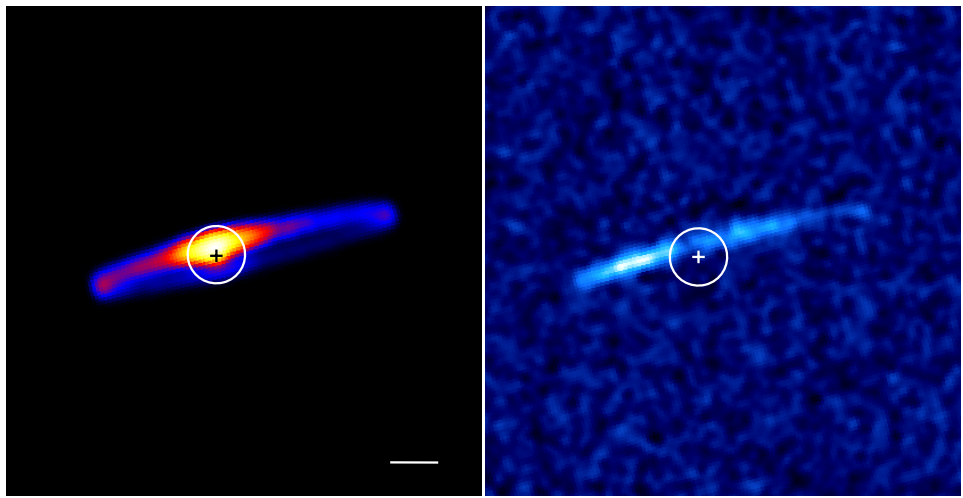}
\caption{Toy radiative transfer model to explore the origin of the east-west brightness asymmetry in HD~106906.  Here we assume an azimuthally uniform belt of scattering grains inclined 5\degr\ from edge-on and with a stellocentric offset equivalent to an eccentricity of 0.2.  The left panel shows total scattered light intensity and the right panel shows polarized intensity. In both panels, the images are convolved with a 3-pixel FWHM Gaussian kernel. Noise was added to the polarized intensity map at approximately the same level as in the observed map to allow for an easy visual comparison. Enhanced forward scattering makes the top of the belt, which resides out of the sky plane, brighter than the bottom of the belt, which is behind the sky plane.  The field of view is $2'' \times 2''$, the scale bar is $0\farcs25$, an $0\farcs125$ radius circle  represents the coronagraphic mask, and the stellar location is marked with a cross.}
\label{fig:toymodel}
\end{figure*}

\begin{figure*}
\centering
\includegraphics[width=3.5in]{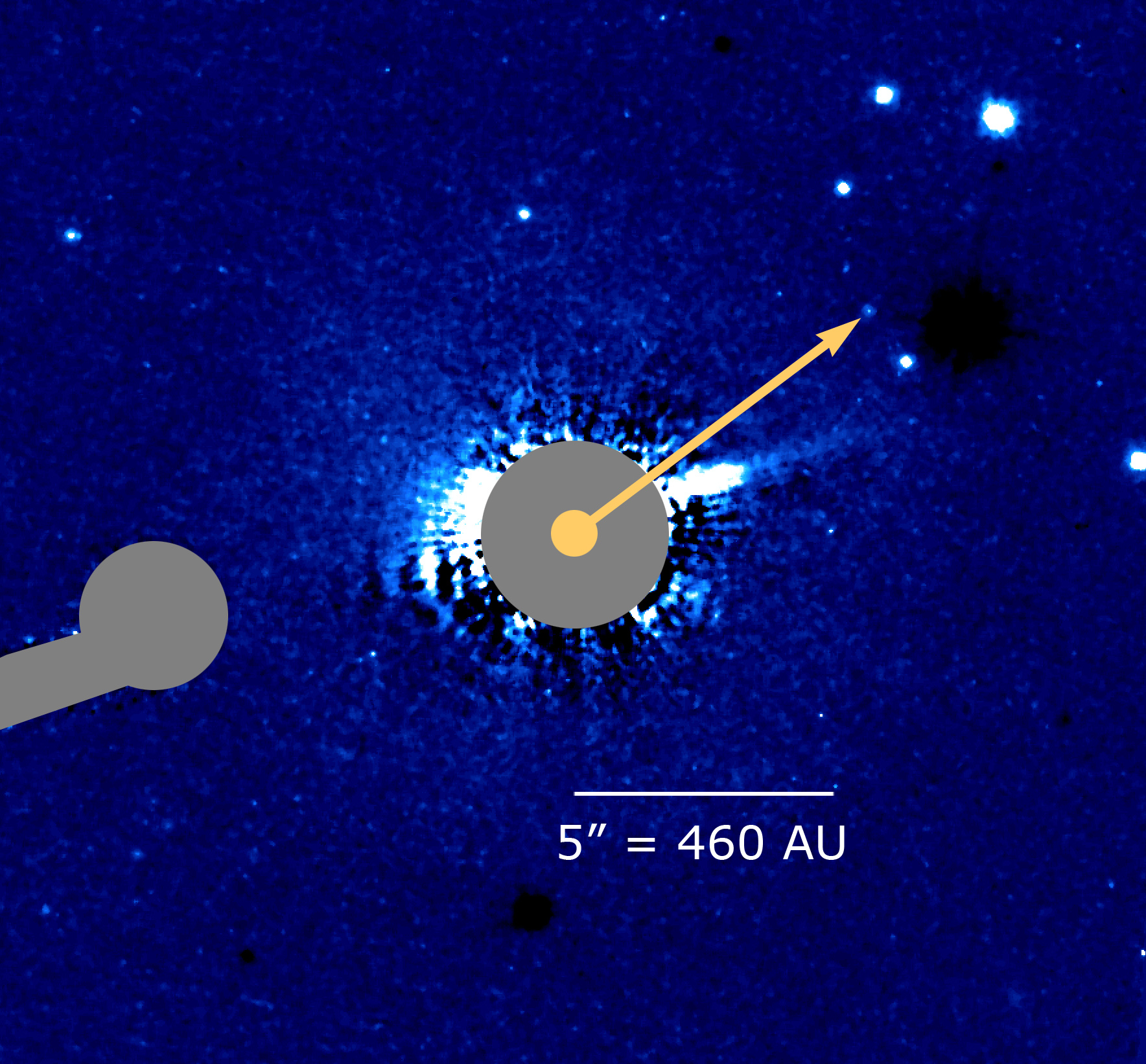}
\caption{HST/ACS/HRC coronagraphic image of HD~106906 in F606W. North is up, east is left.  After PSF subtraction, an extended nebulosity is detected as a sharp feature extending nearly 500 AU to the west (sensitivity limited value). The region east of the star lacks a mirror image of the disk, instead appearing as a diffuse nebulosity spanning PA$\sim 45\degr - 90 \degr$.  The yellow arrow (length 7\farcs14, PA=307\fdg1) points to HD~106906b.}
\label{fig:HST}
\end{figure*}

\begin{figure*}
\centering
\includegraphics[width=3.5in]{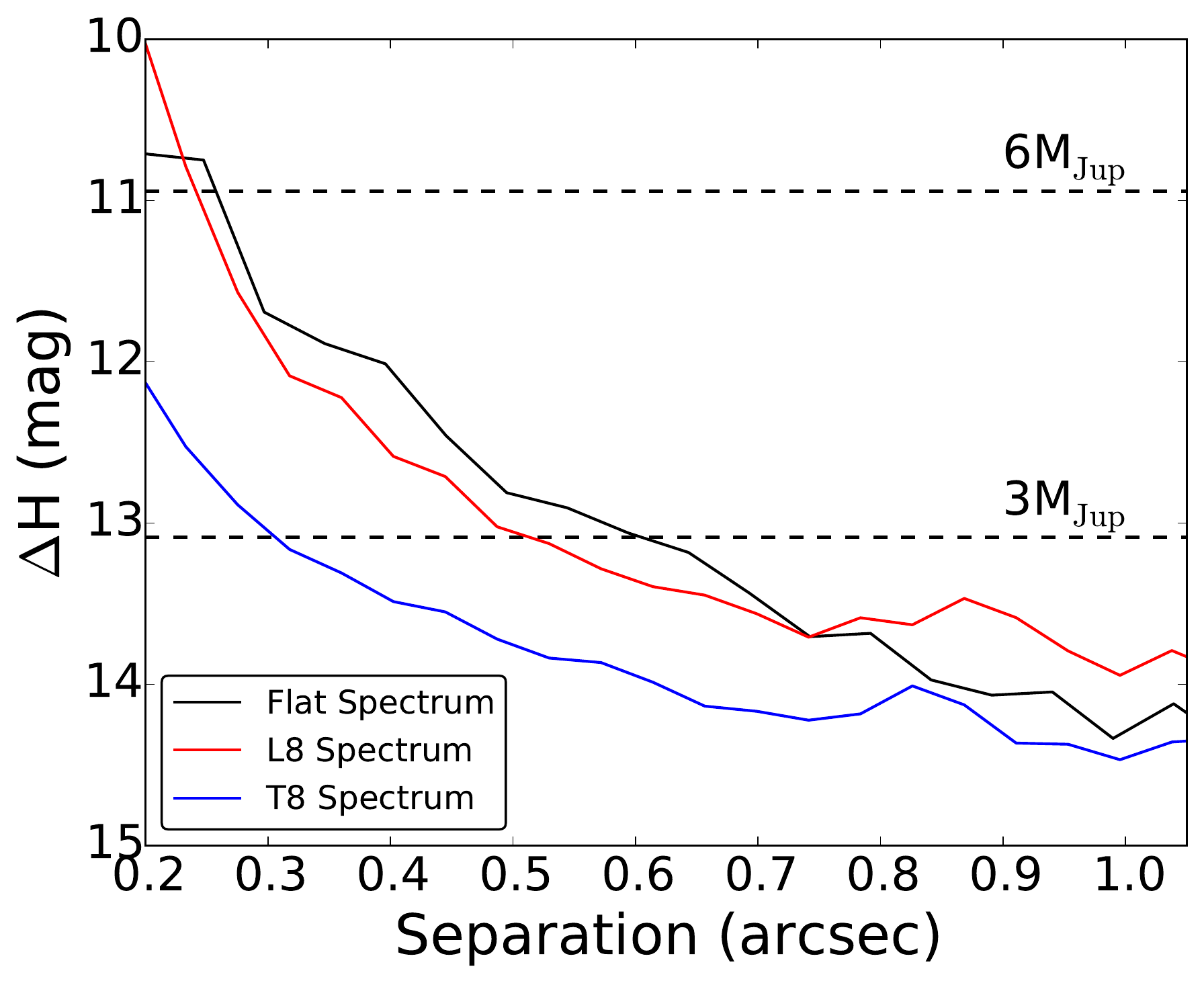}
\includegraphics[width=3.5in]{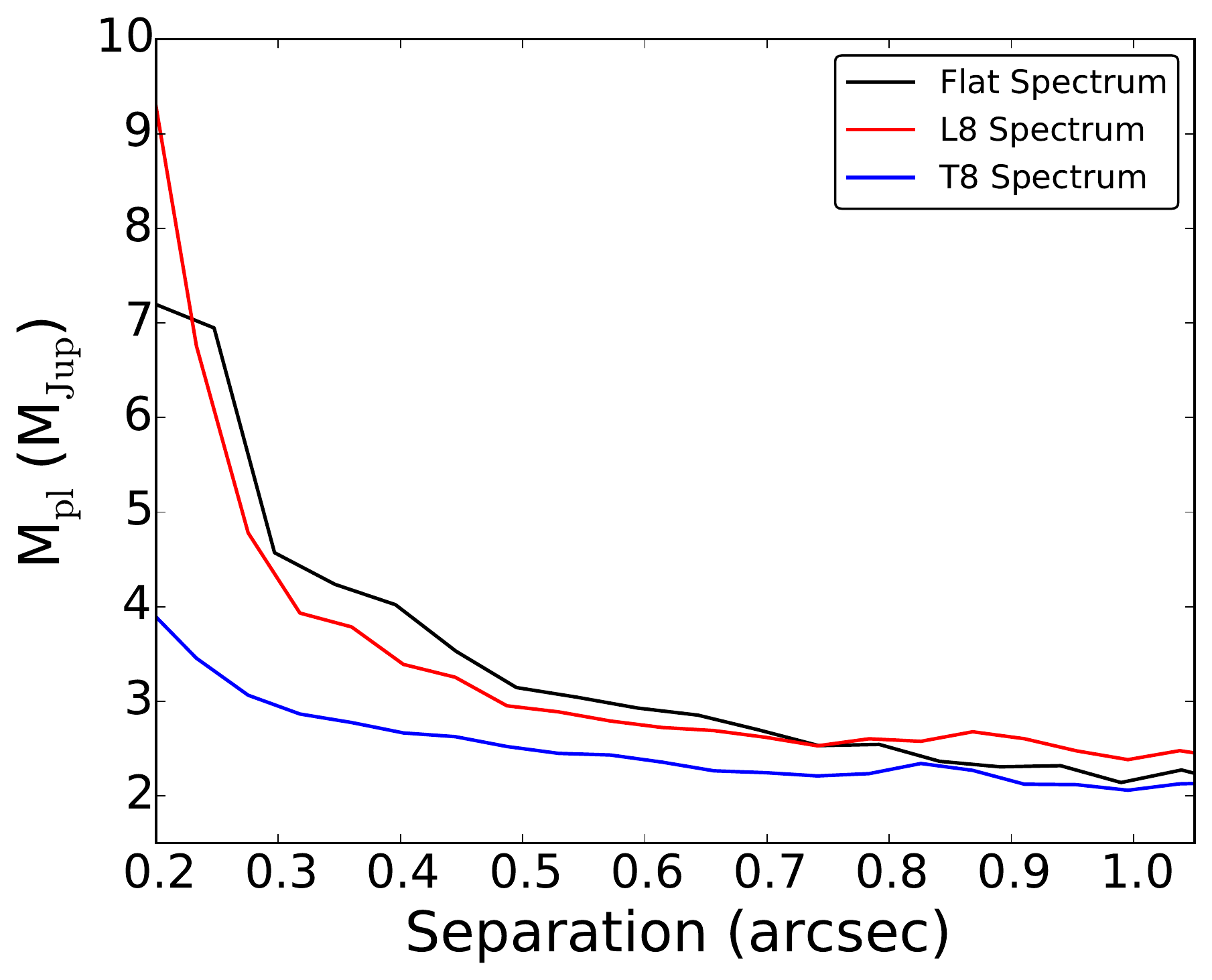}
\caption{{\bf LEFT:} 5-$\sigma$ contrast in the $H$-band as a function of angular separation from the star. Three contrast curves are shown for different companion spectral shapes: flat (black), L8 (red), and T8 (blue). Over-plotted is the detection threshold for a 3M$_{\text{Jup}}$ and 6M$_{\text{Jup}}$ planet calculated using a 13~Myr BT-Settl model \citep{allard12a}. {\bf RIGHT:} Minimum detectable planet mass (5-$\sigma$ limit) as a function of angular separation.}
\label{fig:Contrast}
\end{figure*}

\begin{figure*}
\centering
\includegraphics[width=4.0in]{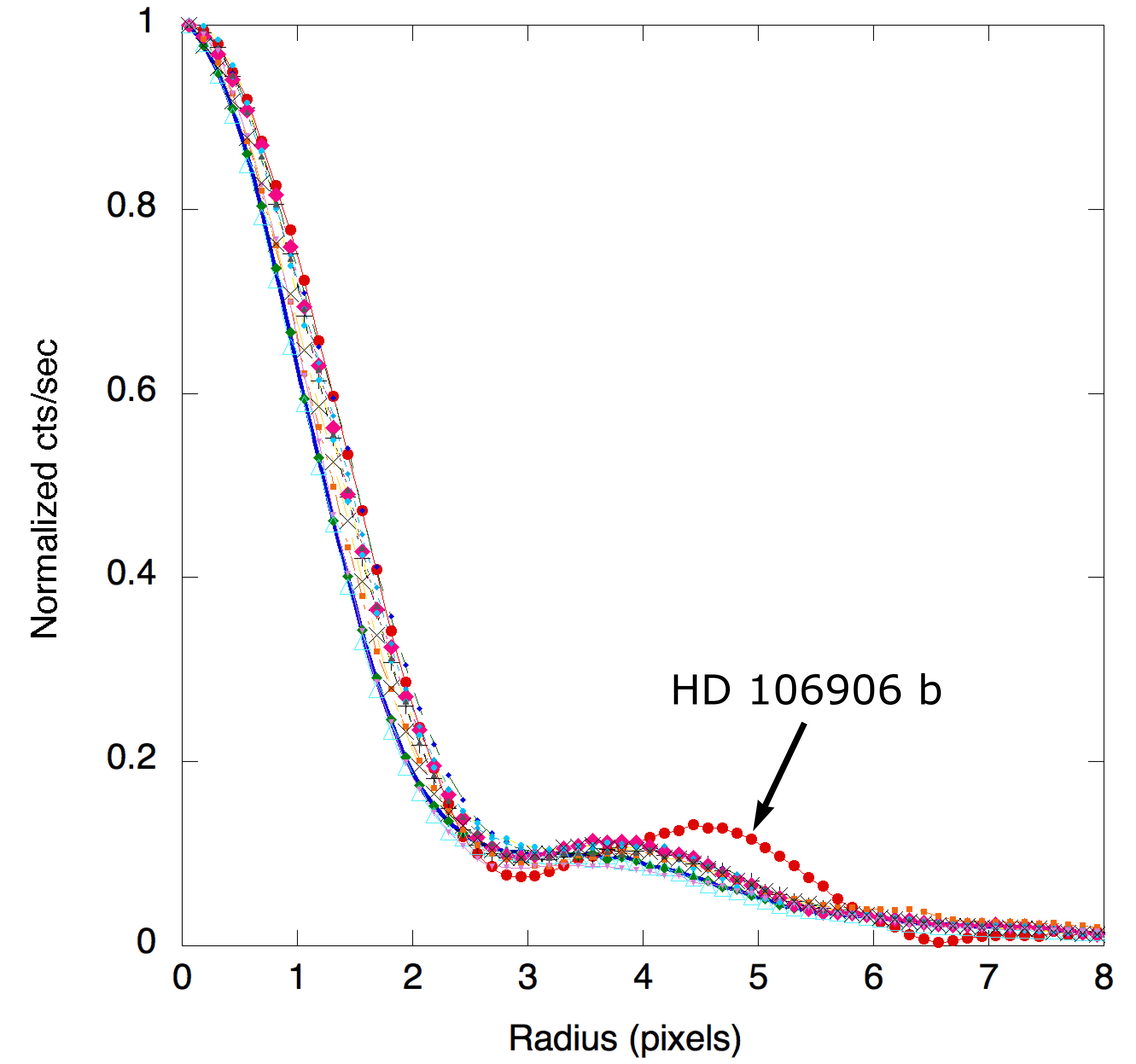}
\caption{Radial intensity profiles (the median value in concentric rings centered on each star) of HD~106906b and 11 other point sources in the HST/HRC F606W field, normalized to unity.  The HRC pixel scale is 25 mas/pixel and the measurements are made with IDP3 \citep{lytle99a}, resampling the image by a factor of eight using bicubic sinc interpolation.  For HD~106906b, the azimuthal morphology of the first Airy ring is similar to the other stars, but there is a distinct outward radial offset in the peak of the first Airy ring, which also contains more flux than the 11 comparison stars.}
\label{fig:RadialProfile}
\end{figure*}

\begin{figure*}
\centering
\includegraphics[width=7in]{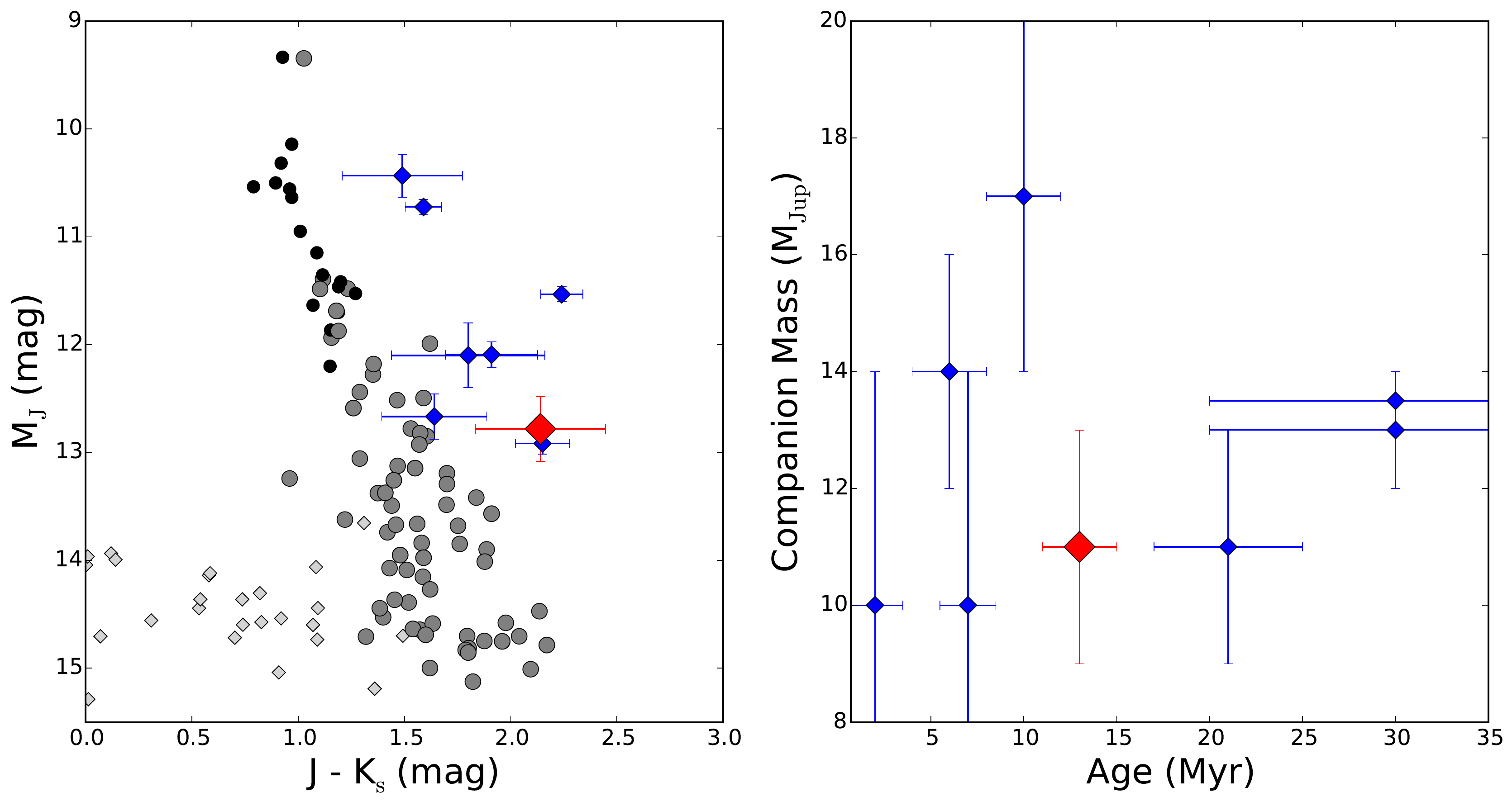}
\caption{{\bf LEFT:} Near-infrared color-magnitude diagram for M dwarfs (black circles), L dwarfs (gray circles), and T dwarfs (gray diamonds) along with a sample of known, young, low-mass companions with ages between 2 and 30 Myr and spectral types from L0 - L4 (blue diamonds).  HD~106906b is marked with a large red diamond.  The M, L, and T dwarf photometry is taken from \citet{dupuy12a}. {\bf RIGHT:} Companion mass as a function of age for the companion sample from the left panel. The sample, presented in Table~\ref{tab:Photometry} and \ref{tab:ModelComp}, includes targets with ages spanning both younger and older than HD106906 with masses comparable to the HD~106906b.}
\label{fig:CMD}
\end{figure*}

\begin{figure*}
\centering
\includegraphics[width=7in]{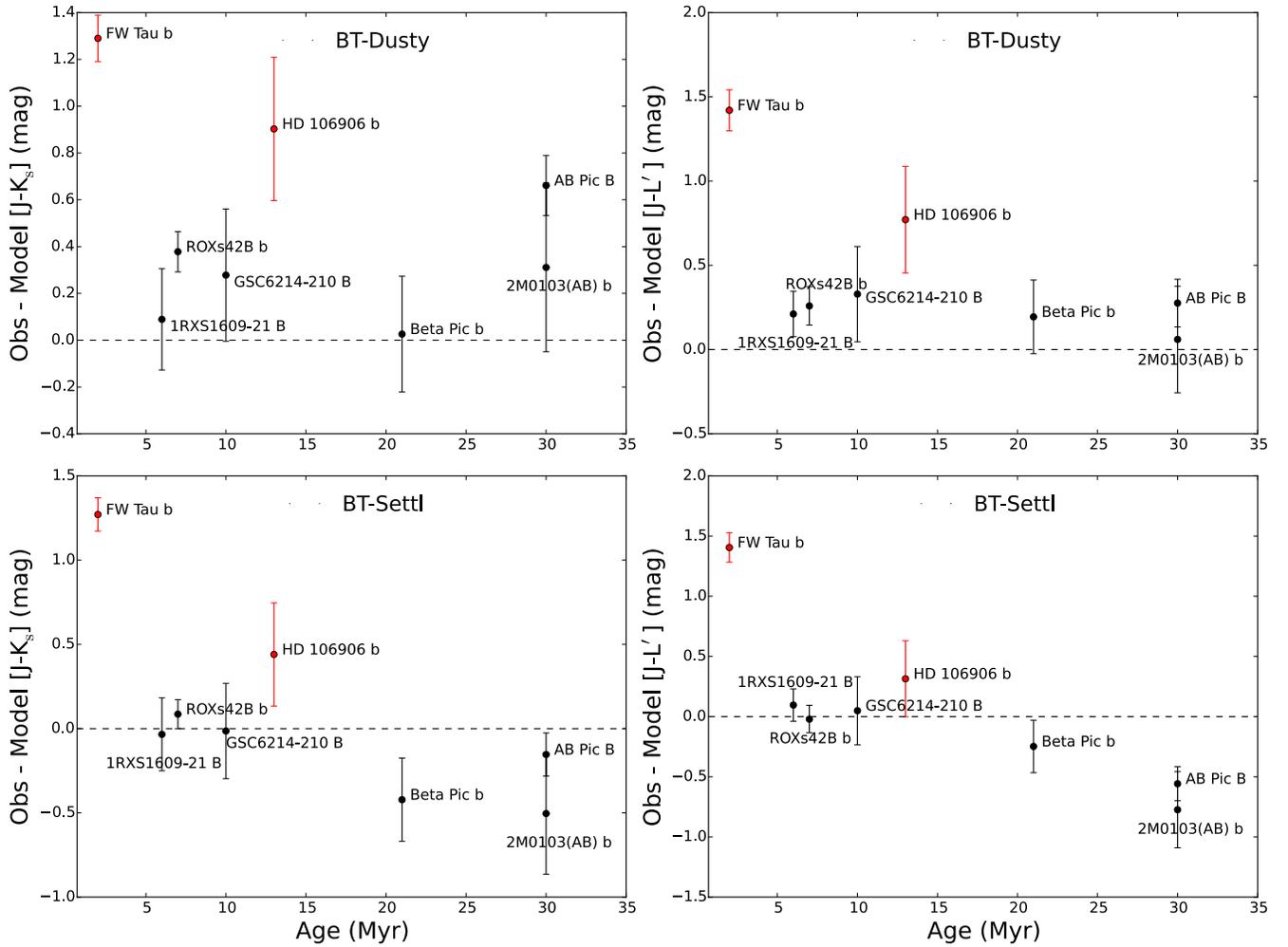}
\caption{The difference between observed and model colors as a function of age for the young companion sample in Table 1.  The dashed line indicates a color difference of zero.  The plot illustrates that only two objects, HD 106906b and FW Tau b, consistently show observed colors redder than the models.}
\label{fig:colorComp}
\end{figure*}

\clearpage

\begin{deluxetable}{|c|l|l|l|l|l|l|l|}
\tabletypesize{\scriptsize} 
\tablecolumns{13}
\tablecaption{Literature photometry for the imaged companions}
\tablehead{Companion & Mass & Age  & \multicolumn{3}{c|}{Apparent Magnitude} & Evidence for  & References \\
            Name & (M$_{\text{Jup}}$) & (Myr) & J (mag) & Ks (mag) & L' (mag) & disk & }
\startdata
HD~106906b   & 11 $\pm$ 2 & 13 $\pm$ 2 & 17.6 $\pm$ 0.3 & 15.46 $\pm$ 0.06 & 14.6 $\pm$ 0.1 &  \emph{HST} extended PSF, optical & [1], [2] \\
 & & & & & & and NIR excess & \\
FW~Taub      & 10 $\pm$ 4 & 2 $\pm$ 1.5 & 17.34 $\pm$ 0.07 & 15.1 $\pm$ 0.1 & 14.3 $\pm$ 0.1 & Acretion signatures \& & [3], [4], [5] \\
 & & & & & & ALMA disk detection & \\
ROXs~42Bb   & 10 $\pm$ 4 & 7 $\pm$ 1.5 & 16.12 $\pm$ 0.07 & 14.53 $\pm$ 0.05 & 13.7 $\pm$ 0.1 & no evidence of disk & [5], [6] \\
GSC~6214-210B  & 17 $\pm$ 3 & 10 $\pm$ 2  & 16.2 $\pm$ 0.2 & 14.8 $\pm$ 0.2 & 13.8 $\pm$ 0.2 &  H$\alpha$, Pa$\beta$, NIR excess & [3], [4], [7], [8] \\
1RXS~1609-2105B & 14 $\pm$ 2 & 6 $\pm$ 2   & 17.9 $\pm$ 0.1 & 15.99 $\pm$ 0.18 & 14.8 $\pm$ 0.1 & Optical and NIR excess & [7], [9], [10] \\
Beta~Picb  & 11 $\pm$ 2 & 21 $\pm$ 4 & 14.11 $\pm$ 0.21 & 12.47 $\pm$ 0.13 & 11.17 $\pm$ 0.06  & none reported & [11], [12], [13], [14] \\
AB~PicB  & 13.5 $\pm$ 0.5 & 30 $\pm$ 10 & 16.3 $\pm$ 0.1 & 14.14 $\pm$ 0.08 & 13.01 $\pm$ 0.09  &  none reported & [15], [16], [17] \\
2M0103(AB)b  & 13 $\pm$ 1 & 30 $\pm$ 10 & 15.5 $\pm$ 0.3 & 13.7 $\pm$ 0.2 & 12.7 $\pm$ 0.1  &  none reported  & [17]
\enddata
\tablenotetext{}{References: [1] \citet{bailey14a}, [2] this paper, [3] \citet{kraus14a}, [4] \citet{bowler14a}, [5] \citet{kraus15a}, [6] \citet{currie14a}, [7] \citet{bailey13a}, [8] \citet{zhou14a}, [9] \citet{lafreniere08a}, [10] \citet{wu15a}, [11] \citet{lagrange09a}, [12] \citet{bonnefoy11}, [13] \citet{binks14a}, [14] \citet{currie11}, [15] \citet{bonnefoy10}, [16] \citet{patience12a}, [17] \citet{delorme13} }
\label{tab:Photometry}
\end{deluxetable}

\begin{deluxetable}{|c|l|ll|l|ll|}
\tabletypesize{\scriptsize} 
\tablecolumns{11}
\tablecaption{Comparison of observed infrared colors vs. model colors for the imaged companions}
\tablehead{Companion & Observed Color & \multicolumn{2}{c|}{Model Color (J-Ks)} & Observed Color & \multicolumn{2}{c|}{Model Color (J-L')} \\
Name & \multicolumn{1}{c|}{J-Ks (mag)} & BT-Settl & BT-Dusty & \multicolumn{1}{c|}{J-L' (mag)} & BT-Settl & BT-Dusty }
\startdata
HD~106906b & 2.14 $\pm$ 0.3 & 1.7 & 1.24 & 3.0 $\pm$ 0.3 & 2.69 & 2.23 \\
FW~Taub & 2.24 $\pm$ 0.1 & 0.97 & 0.95  & 3.09 $\pm$ 0.12 & 1.69 & 1.67 \\
ROXs~42Bb & 1.59 $\pm$ 0.09 & 1.50 & 1.21 & 2.42 $\pm$ 0.11 & 2.44 & 2.16 \\
GSC~6214-210B & 1.49 $\pm$ 0.28 & 1.50 & 1.21 & 2.49 $\pm$ 0.28 & 2.44 & 2.16 \\
1RXS~1609-2105B & 1.91 $\pm$ 0.22 & 1.94 & 1.82 & 3.1 $\pm$ 0.13 & 3.00 & 2.89 \\
Beta~Picb & 1.64 $\pm$ 0.25 & 2.06 & 1.61 & 2.94 $\pm$ 0.22 & 3.19 & 2.75 \\
AB~PicB & 2.15 $\pm$ 0.13 & 2.30 & 1.49 & 3.28 $\pm$ 0.13 & 3.57 & 2.74 \\
2M0103(AB)b & 1.8 $\pm$ 0.36 & 2.30 & 1.49 & 2.8 $\pm$ 0.32 & 3.57 & 2.74 
\enddata
\label{tab:ModelComp}
\end{deluxetable}
\clearpage

\end{document}